\documentclass{article}
\usepackage{amsfonts}
\usepackage{amsmath}

\input{tcilatex}
\begin{document}

\title{Is a Circular Orbit Possible According to General Relativity?}
\author{F.T. Hioe* and David Kuebel \and Department of Physics, St. John
Fisher College, Rochester, NY 14618 \and and \and Department of Physics \&
Astronomy, University of Rochester, Rochester, NY 14627, USA}
\maketitle

\begin{abstract}
A new parameter space is used to classify circular orbits in the
Schwarzschild metric.

\end{abstract}

In Newtonian mechanics, the answer to the question "Is a circular orbit
possible for a particle in a gravitational field?" is a simple yes. In
general relativity, the answer that is given in most texts on this subject
is not based on the exact solution of the orbit in the Schwarzschild metric,
but is based on an analysis that makes use of the effective potential. Our
treatment of this question is based on the explicit analytic expressions for
the trajectories of a particle with a non-zero mass in the Schwarzschild
metric that include bound, unbound and terminating orbits that we presented
recently [1,2]. An understanding of how the various circular orbits given by
the effective potential analysis are related to the exact solutions is
important for clarity and for giving a definitive answer to the question.

As we pointed out in ref.1, the trajectories of a non-zero mass particle can
be conveniently characterized and placed on a "map" in a dimensionless
parameter space ($e,s$), where we have called $e$ the energy parameter and $%
s $ the field parameter. The parameter space ($e,s$), where $0\leq e\leq
\infty $, and $0\leq s\leq \infty $, can be divided into two regions: Region
I characterized by $0\leq s\leq s_{1}$, where $s_{1}(e)$ is well defined,
allows terminating orbits as well as the elliptic- ($0\leq e<1$), parabolic-
($e=1$), and hyperbolic-type ($e>1$) orbits, while Region II characterized
by $s>s_{1}$ allows terminating orbits only. It will be seen that this
parameter space needs to be extended in order to represent all possible
circular orbits.

The Schwarzschild geometry is the static spherically symmetric gravitational
field in the empty space surrounding some massive spherical object (which we
shall refer to as a star or a black hole even though this also includes the
case of a planet such as Jupiter or Earth) of mass $M$. We will consider a
test particle (which can be a spacecraft, an asteroid, or a planet), with
its position relative to the star or black hole described in the spherical
coordinates $r,\theta ,\phi $, moving in the equatorial plane $\theta =\pi
/2 $. Let $E$ be the total energy of the particle in its orbit and $m_{0}$
its rest mass, and define

\begin{equation}
\kappa \equiv \frac{E}{m_{0}c^{2}}.
\end{equation}

Let

\begin{equation}
h\equiv r^{2}\overset{\cdot }{\phi },
\end{equation}

which can be identified as the angular momentum per unit rest mass of the
particle, and where the derivative $\overset{\cdot }{}$ represents $d/d\tau $
and $\tau $ is the proper time along the path. The Schwarzschild radius $%
\alpha $ is defined by

\begin{equation}
\alpha \equiv \frac{2GM}{c^{2}},
\end{equation}

where $G$ is the universal gravitation constant, and $c$ is the speed of
light. Making the substitution

\begin{equation}
u=\frac{1}{r},
\end{equation}

the 'combined' energy equation [3]

\begin{equation}
\overset{\cdot }{r}^{2}+\frac{h^{2}}{r^{2}}\left( 1-\frac{\alpha }{r}\right)
-\frac{c^{2}\alpha }{r}=c^{2}(\kappa ^{2}-1)
\end{equation}

together with eq.(2) can be expressed as

\begin{equation}
\left( \frac{du}{d\phi }\right) ^{2}+u^{2}=\frac{c^{2}(\kappa ^{2}-1)}{h^{2}}%
+\frac{c^{2}\alpha }{h^{2}}u+\alpha u^{3}.
\end{equation}

As shown in ref.1, it is more convenient to work with the dimensionless
parameter $U$ defined by

\begin{equation}
U=\frac{1}{4}\left( \frac{\alpha }{r}-\frac{1}{3}\right) ,
\end{equation}

in terms of which eq.(6) can be expressed as

\begin{equation}
\left( \frac{dU}{d\phi }\right) ^{2}=4U^{3}-g_{2}U-g_{3},
\end{equation}

where

\begin{eqnarray}
g_{2} &=&\frac{1}{12}-s^{2},  \notag \\
g_{3} &=&\frac{1}{216}-\frac{1}{12}s^{2}+\frac{1}{4}(1-e^{2})s^{4},
\end{eqnarray}

and where the dimensionless parameter

\begin{equation}
e=\left[ 1+\frac{h^{2}c^{2}(\kappa ^{2}-1)}{(GM)^{2}}\right] ^{1/2}
\end{equation}

that ranges from $0$ to $\infty $ is called the energy eccentricity
parameter or simply the energy parameter, and the dimensionless parameter

\begin{equation}
s=\frac{GM}{hc}
\end{equation}

that ranges from $0$ to $\infty $ is called the gravitational field
parameter or simply the field parameter. It is seen that the energy
parameter $e$ actually depends on the energy, the gravitational field, and
the angular momentum of the particle, and the field parameter $s$ depends on
the gravitational field and the angular momentum of the particle, but not on
the energy.

The parameter space in our work presented in refs.1 and 2 was confined to $%
e\geq 0$ and $s\geq 0$. General relativity in fact allows an extended
parameter space that includes $e^{2}<0.$ Let us first note what $e^{2}\geq 0$
and $e^{2}<0$ mean physically. From eqs.(2), (5), and (10), we find that $%
e^{2}\geq 0$ or $<0$ respectively imply

\begin{equation*}
\overset{\cdot }{r}^{2}+\left( r\overset{\cdot }{\phi }-\frac{GM}{r^{2}%
\overset{\cdot }{\phi }}\right) ^{2}\geq or<\frac{2GM}{c^{2}}r\overset{\cdot 
}{\phi }^{2}.
\end{equation*}

The right-hand side of the above inequalities is $0$ in the Newtonian limit,
and thus it is important to note that while the parameter space $e^{2}\geq 0$
has its Newtonian correspondence, the "extended" space characterized by $%
e^{2}<0$ does not have a Newtonian correspondence. In the following
discussion, we shall first confine our results to the parameter space $e\geq
0$, and then discuss the additional results from those that include the
extended parameter space.

The discriminant $\Delta $ of the cubic equation

\begin{equation}
4U^{3}-g_{2}U-g_{3}=0
\end{equation}

is defined by

\begin{equation}
\Delta =27g_{3}^{2}-g_{2}^{3}
\end{equation}

and divides the parameter space $(e,s)$ for $e\geq 0$ and $s\geq 0$ into two
regions: Region I corresponding to $\Delta \leq 0$ covers the values of the
field parameter $0\leq s\leq s_{1}$, and Region II corresponding to $\Delta
>0$ covers the values of $s>s_{1}$, where $s_{1}$ as a function of $e$ is
given by [1]

\begin{equation}
s_{1}^{2}=\frac{1-9e^{2}+\sqrt{(1-9e^{2})^{2}+27e^{2}(1-e^{2})^{2}}}{%
27(1-e^{2})^{2}},
\end{equation}

for $e\neq 1$, and $s_{1}^{2}=1/16$ for $e=1$. We note that the parameter
value $s_{1}$ that characterizes the (upper) boundary of Region I decreases
very little from $s_{1}=(2/27)^{1/2}=0.272165..$ for $e=0$ to $%
s_{1}=1/4=0.250000$ for $e=1$, but that it decreases more quickly as $e$
increases beyond $1$ and $s_{1}^{2}\rightarrow (\sqrt{27}e)^{-1}\rightarrow
0 $ as $e\rightarrow \infty $ [4]. Region I has terminating and
non-terminating orbits that include bound (for $0\leq e<1$) and unbound
orbits (for $e\geq 1$), and Region II has terminating orbits only.

In Region I ($\Delta \leq 0$), the three roots of the cubic equation (13)
are all real. We call the three roots $e_{1},e_{2},e_{3}$ and order them so
that $e_{1}>e_{2}>e_{3}$. Define a dimensionless distance $q$ measured in
units of the Schwarzschild radius $\alpha $ by

\begin{equation}
q\equiv \frac{r}{\alpha },
\end{equation}

which is related to $U$ of eq.(7) by

\begin{equation}
\frac{1}{q}=\frac{1}{3}+4U.
\end{equation}

The equation for the non-terminating orbits in Region I, bound and unbound,
and for $0\leq e\leq \infty $, is given in terms of Jacobian elliptic
functions [5] of modulus $k$ by

\begin{equation}
\frac{1}{q}=\frac{1}{3}+4e_{3}+4(e_{2}-e_{3})sn^{2}(\gamma \phi ,k),
\end{equation}

where $e_{1}>e_{2}\geq U>e_{3}$ [6]. The constant $\gamma $ appearing in the
argument and the modulus $k$ of the Jacobian elliptic functions are given in
terms of the three roots of the cubic equation (12) by

\begin{eqnarray}
\gamma &=&(e_{1}-e_{3})^{1/2},  \notag \\
k^{2} &=&\frac{e_{2}-e_{3}}{e_{1}-e_{3}},
\end{eqnarray}

and the roots $e_{1},e_{2},e_{3}$ are given by

\begin{eqnarray}
e_{1} &=&2\left( \frac{g_{2}}{12}\right) ^{1/2}\cos \left( \frac{\theta }{3}%
\right) ,  \notag \\
e_{2} &=&2\left( \frac{g_{2}}{12}\right) ^{1/2}\cos \left( \frac{\theta }{3}+%
\frac{4\pi }{3}\right) ,  \notag \\
e_{3} &=&2\left( \frac{g_{2}}{12}\right) ^{1/2}\cos \left( \frac{\theta }{3}+%
\frac{2\pi }{3}\right) ,
\end{eqnarray}

where

\begin{equation}
\cos \theta =g_{3}\left( \frac{27}{g_{2}^{3}}\right) ^{1/2}.
\end{equation}

The modulus $k$ of the elliptic functions has a range $0\leq k^{2}\leq 1$ as 
$\theta $ varies from $0$ to $\pi $. The period of $sn^{2}(\gamma \phi ,k)$
is $2K(k)$, where $K(k)$ is the complete elliptic integral of the first kind
[5]. For $k=0$, $sn(x,0)=\sin x,$ $cn(x,0)=\cos x,$ $dn(x,0)=1$, and for $%
k^{2}=1$, $sn(\gamma \phi ,1)=\tanh (\gamma \phi )$, $cn(\gamma \phi
,1)=dn(\gamma \phi ,1)=\sec h(\gamma \phi )$. As $k^{2}$ increases from $0$
to $1$, $K(k)$ increases from $\pi /2$ to $\infty $.

The orbits given by eq.(17) for $0\leq e<1$, $e=1$, and $e>1$ will be
referred to as the elliptic-, parabolic- and hyperbolic-type orbits
respectively, even though their shapes can differ greatly from the conic
sections that arise in Newtonian mechanics, as shown in ref.1. They
correspond to the cases for which $0\leq \kappa ^{2}<1$, $\kappa ^{2}=1$,
and $\kappa ^{2}>1$ respectively, where $\kappa $ is defined by eq.(1).

For the elliptic-type ($0\leq e<1$) orbit, the distance $r$ of the planet
from the center of the star or black hole assumes the same value when its
polar angle $\phi $ increases from $\phi $ to $\phi +2K/\gamma $. Comparing
this with the increase of $\phi $ from $\phi $ to $\phi +2\pi $ in one
revolution for the planet, the precession angle is the difference of the two
and is given by

\begin{equation}
\Delta \phi =\frac{2K(k)}{\gamma }-2\pi .
\end{equation}

For $k^{2}$ close to the value $1$, the particle can make many revolutions
around the star or black hole before assuming a distance equal to its
initial distance.

The maximum distance $r_{\max }$ (for $0\leq e\leq 1$) of the particle from
the star or black hole and the minimum distance $r_{\min }$ (for $0\leq
e\leq \infty $) of the particle from the star or black hole, or their
corresponding dimensionless forms $q_{\max }$ $(=r_{\max }/\alpha )$ and $%
q_{\min }$ $(=r_{\min }/\alpha ),$ are obtained from eq.(17) when $\gamma
\phi =0$ and when $\gamma \phi =K(k)$ respectively, and they are given by

\begin{equation}
\frac{1}{q_{\max }}=\frac{1}{3}+4e_{3},
\end{equation}

and

\begin{equation}
\frac{1}{q_{\min }}=\frac{1}{3}+4e_{2}.
\end{equation}

In contrast to the energy eccentricity $e$ defined by eq.(10), the true
eccentricity $\varepsilon $ of the elliptic- and parabolic-type type ($0\leq
e\leq 1$) orbit given by eq.(17) is defined by

\begin{equation}
\varepsilon =\frac{r_{\max }-r_{\min }}{r_{\max }+r_{\min }}=\frac{q_{\max
}-q_{\min }}{q_{\max }+q_{\min }}=\frac{e_{2}-e_{3}}{1/6+e_{2}+e_{3}},
\end{equation}%
and the true eccentricity $\varepsilon $ coincides with the eccentricity $e$
only when $s=0$ for $0\leq e<1$ ($\varepsilon =1$ when $e=1$ for all values
of $s$). The relationship between $e$ and $\varepsilon $ was exhibited in
ref.1.

We now describe in some detail two special cases (i) $k^{2}\simeq 0$ and
(ii) $k^{2}=1$ which relate to the possibility of circular orbits for the
parameter space being considered ($e\geq 0$).

(i) The special case of $k^{2}\simeq 0$.

First we note that if $k^{2}=0$ exactly, then it follows from eqs.(18) and
(19) that this requires $\theta =0$, and thus $e_{1}=2(g_{2}/12)^{1/2}$, $%
e_{2}=e_{3}=-(g_{2}/12)^{1/2}$, and from eq.(20) that $s=0$, from eq.(13)
that $\Delta =0$, and from eq.(24) that $\varepsilon =0$. Thus we find $%
e_{1}=1/6$, $e_{2}=e_{3}=-1/12$, and $\gamma =1/2$. Because $s=0$ implies
zero gravitational field, it is important to recognize that the classical
Newtonian case corresponds to the case of $s\simeq 0$ and $k^{2}\simeq 0$,
and not $s=0$ and $k^{2}=0$ exactly.

Consider the Newtonian limit $s\simeq 0$ and $k^{2}\simeq 0$. The constant $%
c^{2}(\kappa ^{2}-1)$ which is $<0$ for a bound orbit and $\geq 0$ for an
unbound orbit, can be identified with $2E_{0}/m$ in the Newtonian limit,
where $E_{0}$ is the sum of the kinetic and potential energies and is $<0$
for a bound orbit, and is $\geq 0$ for an unbound orbit, and $m$ is the mass
of the particle (which approaches $m_{0}$). In this limit we therefore obtain

\begin{equation*}
e\simeq \left[ 1+\frac{2E_{0}h^{2}}{m(GM)^{2}}\right] ^{1/2},
\end{equation*}

and eq.(17) reduces to the corresponding Newtonian orbits given by

\begin{equation*}
\frac{1}{r}\simeq \frac{GM}{h^{2}}(1-e\cos \phi )
\end{equation*}

for $e>0$, while the case for $e=0$ will be analyzed more carefully below.
We note that, just like the Newtonian orbit equation above for which the
cases $0\leq e<1$ and $e\geq 1$ give rise to bound (elliptic) and unbound
(parabolic and hyperbolic) orbits, the single analytic expression
represented by eq.(17) with the same criteria for $e$ gives rise to
precessing bound (elliptic-type) and unbound (parabolic- and
hyperbolic-type) orbits. We should also note that in general relativity a
second parameter $s$ with a range $0\leq s\leq s_{1}$, where $s_{1}$ is
given by eq.(14), is needed, for these three types of orbits to be realized.
For the case $s>s_{1}$, all orbits are terminating.

We now consider the Newtonian limit characterized by very small values of $s$
and $k^{2}$ for the case $e=0$ more precisely. Setting $e=0$ in $g_{3}$ in
eq.(9) and using eq.(20), the expansion of $\cos \theta $ is found to be

\begin{equation}
\cos \theta =1-2^{2}\cdot 3^{3}s^{6}-2\cdot 3^{5}\cdot 5s^{8}-...
\end{equation}

from which we find

\begin{equation}
\theta =2\cdot 3\sqrt{2\cdot 3}s^{3}\left( 1+\frac{3^{2}\cdot 5}{2^{2}}%
s^{2}+...\right) .
\end{equation}

From eq.(19), we then calculate the expansions [7]

\begin{eqnarray}
e_{1} &=&\frac{1}{6}-s^{2}-3s^{4}-20s^{6}+...  \notag \\
e_{2} &=&-\frac{1}{12}+\frac{1}{2}s^{2}+\frac{\sqrt{2}}{2}s^{3}+\frac{3}{2}%
s^{4}+\frac{21\sqrt{2}}{8}s^{5}+10s^{6}+...  \notag \\
e_{3} &=&-\frac{1}{12}+\frac{1}{2}s^{2}-\frac{\sqrt{2}}{2}s^{3}+\frac{3}{2}%
s^{4}-\frac{21\sqrt{2}}{8}s^{5}+10s^{6}+....
\end{eqnarray}

From eq.(18), we obtain

\begin{eqnarray}
\gamma &=&\frac{1}{2}-\frac{3}{2}s^{2}+\frac{\sqrt{2}}{2}s^{3}-\frac{27}{4}%
s^{4}+\frac{33\sqrt{2}}{8}s^{5}+...  \notag \\
k^{2} &=&4\sqrt{2}s^{3}\left( 1+\frac{45}{4}s^{2}+...\right) ,
\end{eqnarray}

and from [5] $K(k)=(\pi /2)(1+k^{2}/4+...)$, we find

\begin{equation}
K(k)=\frac{\pi }{2}\left( 1+\sqrt{2}s^{3}+...\right) .
\end{equation}

Using the above equations and eq.(19), the orbit equation (17) becomes

\begin{equation*}
\frac{1}{q}\simeq 2s^{2}\left\{ 1-\sqrt{2}s\left[ 1-2sn^{2}(\gamma \phi ,k)%
\right] \right\} ,
\end{equation*}

or, after substituting eqs.(3), (11) and (15),

\begin{equation}
\frac{1}{r}\simeq \frac{GM}{h^{2}}\left\{ 1-\sqrt{2}s[1-2sn^{2}(\gamma \phi
,k)]\right\} ,
\end{equation}

where $\gamma $ and $k^{2}$ are given by eq.(28). The elliptic function $%
sn(\gamma \phi ,k)$ can be expanded in power series in $s$ for $e=0$ as

\begin{equation*}
sn(\gamma \phi ,k)=\sin \frac{\phi }{2}-\frac{3\phi }{2}s^{2}\cos \frac{\phi 
}{2}+...
\end{equation*}

To order $s$, we find that the Newtonian limit of the orbit equation for $%
e=0 $ can be expressed as

\begin{equation}
\frac{1}{r}=\frac{GM}{h^{2}}\left( 1-\sqrt{2}s\cos [(1-\delta )\phi ]\right)
,
\end{equation}

where $\delta \simeq 3s^{2}$. Thus, according to general relativity, the
Newtonian limit (for $s$ and $k^{2}$ very small) of the orbit of a particle
for $e=0$ is not a true circular orbit but is an orbit with an eccentricity
equal to $\sqrt{2}s\equiv \sqrt{2}GM/(hc)$, and is thus an elliptical orbit.

Using eqs.(24) and (27), we find the true eccentricity is given by

\begin{equation*}
\varepsilon =\sqrt{2}s+...
\end{equation*}

in agreement with that given in eq.(31). Only if $s=0$ does the orbit
represented by eq.(31) become a true circular orbit, but then it is the case
of zero gravitational field and the radius $h^{2}/(GM)$ of the orbit is
infinite ($h^{2}/(GM)=(hc/GM)(h/c)=(\infty )(h/c)$ because $s\equiv
GM/(hc)=0 $), and thus the particle is going in a straight line.

The precessing elliptical orbit for small $s$ is represented by eq.(31) or
eq.(30). The angle of precession, from eq.(21), is given by

\begin{equation}
\Delta \phi \simeq 6\pi s^{2}=\frac{6\pi (GM)^{2}}{(hc)^{2}}.
\end{equation}

We recall from previous work [1, 2, 7] that for $e>0$, $\gamma
=[1-(3-e)s^{2}+...]$, $k^{2}=4es^{2}+...$, the approximate precessing
elliptical orbit equation is $1/r=(GM/h^{2})\{1-e\cos [(1-\delta )\phi ]\}$
or $1/r=(GM/h^{2})\{1-e[1-2sn^{2}(\gamma \phi ,k)]\}$. Note that to the
order of $s^{2}$, we get the same precession angle $\Delta \phi $ given by
eq.(32) for all elliptical orbits independent of $e$.

We now consider the second special case for $k^{2}$.

(ii) The special case of $k^{2}=1$.

It follows from eqs.(18)-(20) and (13) that $\cos \theta =-1$, $e_{1}=e_{2}=%
\sqrt{g_{2}/12}$, $e_{3}=-\sqrt{g_{2}/3}$, $\gamma =(3g_{2}/4)^{1/4}$, $%
\Delta =0$. The range of $s$ values is given by eq.(14) and gives the
boundary between Regions I and II. On this boundary defined by $k^{2}=1$ we
have $g_{2}>0$, $g_{3}<0$, and the relation between the two is $\sqrt[3]{%
g_{3}}=-\sqrt{g_{2}/3}$.

For $k^{2}=1$, eq.(17) becomes

\begin{equation}
\frac{1}{q}=\frac{1}{3}+4\sqrt{\frac{g_{2}}{12}}-12\sqrt{\frac{g_{2}}{12}}%
\sec h^{2}(\gamma \phi ,k).
\end{equation}

The orbit will be referred to as an asymptotic orbit. For $0\leq e\leq 1$,
the particle starts from an initial position $q_{\max }$ at $\phi =0$ and
ends up at $\phi =\infty $ circling the star or black hole with a radius
that asymptotically approaches $q_{\min }$, where $q_{\min }$ is given by

\begin{equation}
\frac{1}{q_{\min }}=\frac{1}{3}+4\sqrt{\frac{g_{2}}{12}},
\end{equation}

where $g_{2}$ is evaluated using $s=s_{1}$ given by eq.(14). For $e>1$, the
particle starts from an initial position at infinity at an angle $\Psi _{1}$
to the horizontal axis given by

\begin{equation*}
\tanh ^{2}(\gamma \Psi _{1})=\frac{2}{3}-\frac{1}{18}\sqrt{\frac{3}{g_{2}}},
\end{equation*}

and ends up at $\phi =\infty $ circling the star or black hole with a radius
that asymptotically approaches $q_{\min }$.

A possible case of circular orbits is the case of those limiting circles
that the asymptotic orbits given by eq.(33) become. The range of the radius $%
q_{c}$ given by $q_{\min }$ of eq.(34), using the values of $s_{1}$ given
after eq.(14), is between $2.25$ for $e=0$ and $1.5$ for $e=\infty $, i.e.
the radii $r_{c}$ of the limiting circular orbits are in the range

\begin{equation}
\frac{3GM}{c^{2}}\leq r_{c}\leq \frac{4.5GM}{c^{2}},
\end{equation}

and they occur for the values of $s=s_{1}$ given by eq.(14). However, it is
clear that these circular orbits are unstable because orbits exist which are
spirally asymptotic to them [8].

Another possible example of a circular orbit comes up as a special case of
the terminating orbits in Region I. The terminating orbit in Region I is
given by [1]

\begin{equation}
\frac{1}{q}=\frac{1}{3}+4\frac{e_{1}-e_{2}sn^{2}(\gamma \phi ,k)}{%
cn^{2}(\gamma \phi ,k)}.
\end{equation}

where $U>e_{1}>e_{2}>e_{3}$ and $\gamma ,k,e_{1},e_{2},e_{3}$ are given by
eqs.(18) and (19). For $0\leq e\leq 1$, the particle, starting from the
polar angle $\phi =0$ in a direction perpendicular to the line joining it to
the star or black hole, at a distance $q_{1}$ from the star or black hole,
plunges into the center of the black hole when its polar angle reaches $\phi
_{1}$, where $q_{1}$ and $\phi _{1}$ are given by

\begin{equation}
\frac{1}{q_{1}}=\frac{1}{3}+4e_{1},
\end{equation}

and

\begin{equation*}
\phi _{1}=\frac{K(k)}{\gamma }.
\end{equation*}

For $e>1$, the particle starts from infinity at a polar angle $\phi =$ $\Psi
_{1}>0$ to the horizontal axis given by

\begin{equation*}
sn^{2}(\gamma \Psi _{1})=-\frac{\frac{1}{3}+4e_{3}}{4(e_{2}-e_{3})},
\end{equation*}

and plunges into the center of the black hole when its polar angle reaches $%
\phi _{1}$.

The terminating orbit given by eq.(36) becomes a circular orbit when $%
k^{2}=1 $, because for $k^{2}=1$, the two roots $e_{1}$ and $e_{2}$ become
equal (see the description before eq.(33)) and $1-\tanh ^{2}(\gamma \phi
)=\sec h^{2}(\gamma \phi )$. We find that the radius $q_{c}$ of the circular
orbit is given by

\begin{equation}
\frac{1}{q_{c}}=\frac{1}{3}+4e_{1}=\frac{1}{3}+4\sqrt{\frac{g_{2}}{12}},
\end{equation}

which is the same as the radius of the limiting circular orbit given by
eq.(34). Thus $q_{c}$ ranges from $2.25$ for $e=0$ to $2$ for $e=1$ and to $%
1.5$ for $e=\infty $, and the radius $r_{c}$ of the circular orbit is in the
same range as that given by eq.(35). However, it is obvious that these
circular orbits are unstable because a small perturbation would make them
into terminal orbits.

We now consider the relationship of these results from the exact solutions
given by eqs.(17) and (36), in particular eqs.(22), (23), (34) and (38), to
the solutions for the circular orbits obtained from the analysis using the
effective potential [9], or from the simple consideration of $r=const.$.
Differentiating eq.(6) with respect to $\phi $, we obtain

\begin{equation}
\frac{d^{2}u}{d\phi ^{2}}+u=\frac{c^{2}\alpha }{2h^{2}}+\frac{3\alpha }{2}%
u^{2}.
\end{equation}

Setting $d^{2}u/d\phi ^{2}=0$ gives a quadratic equation in $u$ and the two
solutions for $u$ can be expressed in $q$ given by eq.(15) and in $g_{2}$
given by eq.(9) as

\begin{equation}
\frac{1}{q}=\frac{1}{3}\pm 4\left( \frac{g_{2}}{12}\right) ^{1/2}.
\end{equation}

The effective potential analysis [9] chose what seemed to be a reasonable
range

\begin{equation}
0\leq s^{2}\leq \frac{1}{12}
\end{equation}

and thus

\begin{equation}
\frac{1}{12}\geq g_{2}\geq 0
\end{equation}

for the solutions given by eq.(40) to remain real. Applying the inequalities
given by eq.(42) to the solution given by eq.(40) with the lower sign
(corresponding to the minimum of the effective potential) gives a range $%
3\leq q\leq \infty $ or a range for the radius $r$ of a possible stable
circular orbit as

\begin{equation}
\frac{6GM}{c^{2}}\leq r\leq \infty ,
\end{equation}

and applying the same inequalities given by eq.(42) to the solution given by
eq.(40) with the upper sign (corresponding to the maximum of the effective
potential) gives a range $1.5\leq q\leq 3$ or a range for the radius $r$ of
a possible unstable circular orbit as\ 

\begin{equation}
\frac{3GM}{c^{2}}\leq r\leq \frac{6GM}{c^{2}}.
\end{equation}

We now want to see what our analytic solutions give for the parameter space
we are using.

First we realize that eq.(17) represents a circular orbit only when $%
e_{2}=e_{3}$. Because $e_{3}$ is negative for the entire range $0\leq
k^{2}\leq 1$ corresponding to $0\leq \theta \leq \pi $ (see eq.(19)), $q$
must be greater than $3$ and hence there is a lower bound to the radius $%
r>6GM/c^{2}$ which is in agreement with eq.(43). However, equating the
expressions for $e_{2}$ and $e_{3}$ given in eq. (19) requires that $\theta
=0$ if we confine ourselves to $e\geq 0$ so that we have $e_{2}$ \ = $e_{3}=$
$-(g_{2}/12)^{1/2}.$ Under these conditions, one has, from eqs.(22) and (23),

\begin{equation}
\frac{1}{q_{\min }}=\frac{1}{q_{\max }}=\frac{1}{3}+4e_{2}=\frac{1}{3}%
-4\left( \frac{g_{2}}{12}\right) ^{1/2},
\end{equation}

which can be identified with the solution (40) with the lower sign. However,
setting $e_{2}=e_{3}$ in eq.(18) shows that the solution given by eq.(45)
requires $k^{2}=0$ which, together with $\theta =0$ also requires $s=0$ as
the only possible value for $s$ (which gives $r=\infty $) (see the
discussion of the special case of $k^{2}\simeq 0$ after eq.(24) and the
discussion after eq.(31)). Thus our exact solution excludes any finite value
of $r$ given in eq.(43) as a possible radius for a circular orbit for the
parameter space we are using. But $s=0$ corresponds to the case of zero
gravitational field which makes this solution useless. Our solution given by
eq.(30) gives the best possible nearly circular orbit; it states that a
stable most "circular" orbit with $e=0$ and with $s\simeq 0$ and $%
k^{2}\simeq 0$ must have a small but nonzero true eccentricity $\varepsilon $
equal to $\sqrt{2}s$ and is thus elliptical and precesses with a precession
angle $\Delta \phi \simeq 6\pi s^{2}$.

We now consider the solution given by the upper sign in eq.(40). It can be
identified with our analytic solutions (17) and (36) when $k^{2}=1$ (see
eqs.(34) and (38)). Note that these are two separate cases: Eq.(34) gives
the limiting radius of an asymptotic orbit from solution (17), and eq.(38)
gives the radius of a circular orbit that a terminating orbit given by
solution (36) becomes under a special case. As previously discussed, our
exact solutions clearly show that both these cases correspond to unstable
circular orbits. Our exact analytic solutions that apply for $\Delta \leq 0$
show the range given by eq.(35) to be the possible range of radii for an
unstable circular orbit for the parameter space we are using.

Thus comparing our analytic solutions while confining ourselves to the
parameter space $e\geq 0$ with those obtained from the effective potential
theory, we find a stable circular orbit only for $r=\infty $ compared with
eq.(43), and an unstable circular orbit in the range given by eq.(35)
compared to that given by eq.(44).

We now consider the extended space characterized by $e^{2}<0$. Our analytic
solution (17) still applies in this region given by $\Delta \leq 0$. The
upper boundary of this region is still given by eq.(14) from the coordinate
point $e^{2}=0$, $s^{2}=2/27$ to the point $e^{2}=-1/3$, $s^{2}=1/12$, which
we shall call the vertex point, at which $g_{2}=g_{3}=0$, and the three
roots of the cubic equation (12) are equal to zero. The upper boundary
characterized by $k^{2}=1$ in the extended parameter space is still defined
by $s_{1}$ (eq.(14) with $2/27<s_{1}^{2}\leq 1/12$) and eqs.(34) and (38)
still hold. Thus in the extended space we have unstable circular orbits with
radii in the range given by

\begin{equation}
\frac{4.5GM}{c^{2}}<r\leq \frac{6GM}{c^{2}},
\end{equation}

which, together with the possible unstable circular orbits with radii given
by eq.(35) in the parameter space $e\geq 0$, make up the total interval
given by eq.(44) which was derived by the effective potential theory. A
similar analysis holds for the range of radii of the limiting circles of the
asymptotic orbits.

The left boundary of Region I is given by $s=s_{1}^{\prime }$, where $%
s_{1}^{\prime }$ is given by

\begin{equation}
s_{1}^{\prime 2}=\frac{1-9e^{2}-\sqrt{(1-9e^{2})^{2}+27e^{2}(1-e^{2})^{2}}}{%
27(1-e^{2})^{2}},
\end{equation}

and it extends from the vertex at $(e^{2},s^{2})=(-1/3,1/12)$ to the origin
at $(e^{2},s^{2})=(0,0)$. It can be verified that along this left boundary
of Region I, $\cos \theta $ given by eq.(20) is equal to $1$, and we have $%
e_{2}=e_{3}$, $k^{2}=0$, and $\varepsilon =0$ from eq.(24). This means that
we have possible stable circular orbits along the left boundary curve with
radii in the range given by eq.(43). It is to be noted that all possible
stable circular orbits with finite radius are in the extended parameter
space $e^{2}<0$, and the physical implication of this extended parameter
space should be noted.

In summary, the analysis of our exact analytic solutions for a non-zero mass
particle in a gravitational field in the Schwarzschild geometry shows the
following results:

(1) A stable circular orbit is not possible in the parameter space $e\geq 0$%
. The most "circular" orbit of a particle in a gravitational field must have
a small but non-zero eccentricity and is thus an elliptical orbit and it
precesses. For $e=0$ and small values of $s$ and $k^{2}$, the orbit equation
is given by eq.(30). However, in the parameter space $e^{2}<0$, a stable
circular orbit with a radius given by eq.(45), where $s=s_{1}^{\prime }$,
and with a range given by $6GM/c^{2}\leq r\leq \infty $, is possible.

(2) An unstable circular orbit with a radius given by eq.(38), where $%
s=s_{1} $, and with a range given by $3GM/c^{2}\leq r\leq 4.5GM/c^{2}$, is
possible in the parameter space $e\geq 0$, and with a range given by $%
4.5GM/c^{2}<r\leq 6GM/c^{2}$ in the parameter space $e^{2}<0$, .

We \ will present a more complete analysis in future work.

References

*Electronic address: fhioe@sjfc.edu

[1] F.T. Hioe and D. Kuebel, Phys. Rev. D 81, 084017 (2010).

[2] F.T. Hioe and D. Kuebel, arXiv:1008.1964 v1 (2010).

[3] M.P. Hobson, G. Efstathiou and A.N. Lasenby: General Relativity,
Cambridge University Press, 2006, Chapters 9 and 10.

[4] The misprint $s_{1}\rightarrow (\sqrt{27}e)^{-1}$ on the 5th line of
Section V in ref.1 should be corrected to $s_{1}^{2}\rightarrow (\sqrt{27}%
e)^{-1}$.

[5] P.F. Byrd and M.D. Friedman: Handbook of Elliptic Integrals for
Engineers and Scientists, 2nd Edition, Springer-Verlag, New York, 1971.

[6] Equations similar to eq.(17) in somewhat different forms were given by
A.R. Forsyth, Proc. Roy. Soc. Lond. A97, 145 (1920), C. Darwin, Proc. Roy.
Soc. Lond. A249, 180 (1958), ibid. A263, 39 (1961) and S. Chandrasekhar: The
Mathematical Theory of Black Holes, Clarendon Press, Oxford 1992, p.
106-109. \ \ \ \ 

[7] The coefficients beyond $s^{2}$ in $e_{2}$ and $e_{3}$ in eq.(20) in
F.T. Hioe, Phys. Lett. A 373, 1506 (2009) are incorrect. However those
errors did not affect any other results presented in that paper.

[8] These orbits were given in E.T. Whittaker: Analytical Dynamics, 4th
Edition, Dover Publications 1937, p.393.

[9] See e. g. \ ref. 3 \ Sections 9.8 and 9.9, also J.V. Narlikar: An
Introduction to Relativity, Cambridge University Press, 2010, Section 9.3.

\ 

\end{document}